\documentclass[a4paper]{article}
\usepackage[english]{babel}
\usepackage[T1]{fontenc}
\usepackage{amsmath}
\usepackage{amssymb}
\usepackage{amsfonts}
\usepackage{amsthm}
\usepackage{url}
\usepackage{graphicx}
\usepackage{geometry}
\begin{document}

\huge{ \textbf{
\begin{flushleft}
Dimensionless Fluctuation Balance Principle: 
New Statistical Perspectives Applied to
Boltzmann, Planck, Fermi-Dirac, Bose-Einstein
and Schrödinger Distributions
\end{flushleft}
} }

\large{ 
\begin{flushleft}
\textbf{Marceliano Oliveira${}^1$, George Valadares${}^2$,
Francisco Rodrigues${}^3$, Marcio Freire${}^4$ }\\
1 UEA Brazil, 2 UFAC Brazil,
3 UVA Brazil, 4 UFC Brazil \\
1 corresponding author - marcelianooliveira@gmail.com
\end{flushleft}
}

\noindent \hrulefill
\textbf{ \large{ \flushleft ABSTRACT } }

\noindent In this work we propose a completely new way to obtain statistics distributions from fluctuations balance. By dimensionless fluctuation analysis we obtain Boltzmann, Planck, Fermi-Dirac, Bose-Einstein and Schrödinger Distributions using the same fundamental principle. Our result point to a general foundation that was successful verified to principal Physics Distributions. We name it as Dimensionless Fluctuation Balance Principle. This is a great achievement which enable us to discuss exchange between different physical quantities, like we do when treat energy conservation when some type of energy is converted to another, but with more generality, because we can exchange one physical quantity to any other. All physics model which needs distribution can take advantage of methodology presented in this paper including: Statistical Physics, Schrödinger's Quantum Mechanics, Nanomaterials, Thin Films and New Materials Modeling. \textbf{Keywords: Fluctuations, PDEs, Boltzmann, Planck, Entropy, Fermi-Dirac, Bose-Einstein, Schrödinger, Distributions.}

\noindent \hrulefill
\thispagestyle{empty}
 
\section{Introduction}

Many times, in Physics we need to extend some fundamental law and
their application \cite{Maxwell1864}, or just produce new models relating
experimental facts slightly different from original \cite{Whestphal2021},
like Perturbation Theory \cite{Merlin2021}, Energy Conservation \cite{Landau1980},
Gauss Divergence Theorem \cite{Basil2011}, Stokes Curl Theorem \cite{Arfken2012},
Reynolds Theorem \cite{Reynolds1902}, each one with their better applicability
to singular context.

Observing some new materials applications \cite{Abayev2006}, we can see
that distributions plays a central role to modeling this system \cite{Pierantoni2021}.
Sometimes some modification in theoretical model due to experimental
requirement is needed as theoretical proposition \cite{Vollath2019}.
We feel the urgency to some principle to obtain distributions quickly,
as the main goal of our work. We name it as \textbf{Dimensionless
Fluctuation Balance Principle}.

Our ``\textit{Ansatz}'' starts when we observe Boltzmann distribution
\cite{Rowlinson2005}, which was applied to Plank's hypothesis resulting
their Radiation Law \cite{Agudelo2010}. In the ultraviolet catastrophe
case as in \cite{Rayleigh1900}, with $\left(E=kT\right)$\footnote{where $k$ is the Boltzmann's constant},
we can see a pair of quantities energy and temperature $\left(E,T\right)$,
and this one takes a form of Boltzmann distribution like $u\left(E,T\right)\propto e^{-E/kT}$
\cite{Trevena2010}.

Our research starts from Dimensionless Fluctuations of $\left(E,T\right)$
pair and we verify how to obtain a Partial Differential Equation (PDE),
which the solution is desired distribution. In this case, Boltzmann
distribution $u\left(E,T\right)\propto e^{-E/kT}$, after this we
try to apply the same procedures to other distributions searching
for a fundamental principle, our tests results was verified
with success to principal Physics Distributions.

This work shows in completely new way, how to obtain statics distributions
from fluctuations balance, theses fluctuations relate fractions from
domain dimensionless terms. A special point about our analysis is
that with this one we can obtain all fundamentals distributions (including
Boltzmann Entropy Law) without probability as a required initial concept.
This achievement doesn't neglect statistical theory in measurement
of states, because we can think any mean value as inner product between
quantity to be measured as distribution on Hilbert Spaces.

\section{Small Fluctuations Balance}

In many cases, when we study a control volume conservation by Reynolds
Transport Theorem \cite{Reynolds1902}, we just observe net rate exchange
of property per unit time. As a classical example, the continuity
equation case when balancing charges in some region of space which
can decrease when some current leaves the volume across boundary surface
\cite{Marion1980}.

Our effort here is to study localized phenomenon, as simple case,
gas inside a closed box or some portion of fluid inside a thermal-mechanical
set. To reach this goal, instead of work with classical balance like
continuity equation, we will study small portions of a property inside
a fraction of material.

As illustrative case study, being a given group of molecules with
the same characteristics, the summation over $N_{i}$\footnote{$N_{i}$ also represent a group of molecules of gas.}
gives the total number of molecules $N$, then


\begin{figure}[ht]

\begin{centering}

\begin{tabular}{ccc}
\includegraphics[scale=0.30]{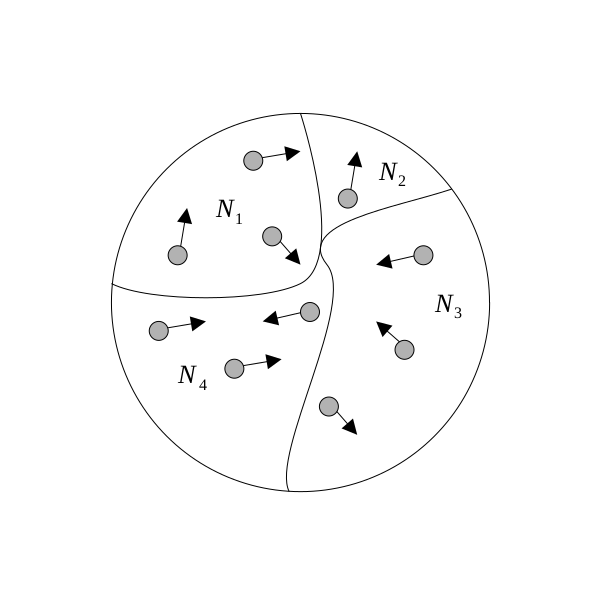} & \includegraphics[scale=0.30]{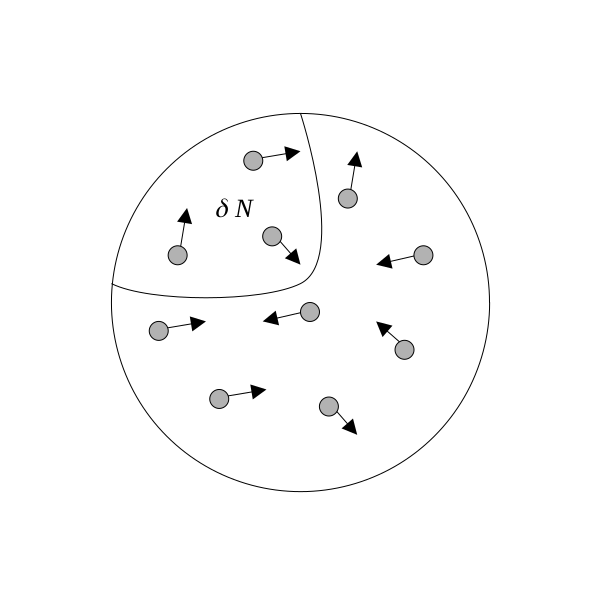} & \includegraphics[scale=0.30]{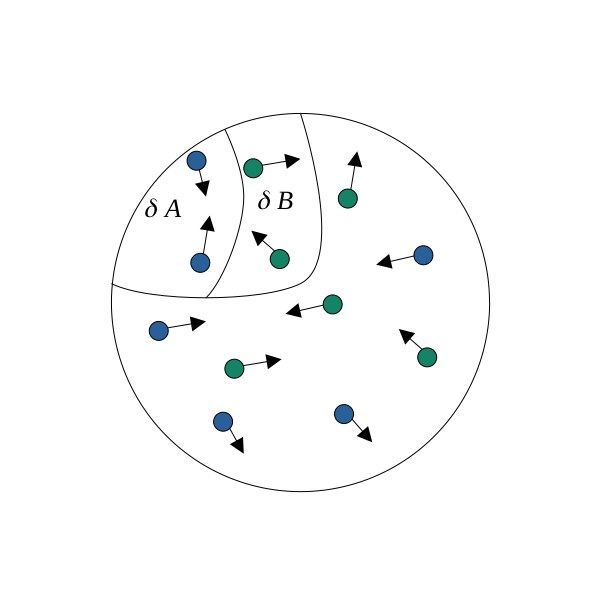}\tabularnewline
\textbf{(a)} & \textbf{(b)} & \textbf{(c)}\tabularnewline
\end{tabular}

\end{centering}

\caption{ 
(a) Representation of molecules arrangement like cluster, fractioning domain in many partitions inspire equation (\ref{eq:1}) which treat about conservation the number of molecules distributed in partitions of cluster. (b) If the cluster is large with many particles, and divided in many partitions, a single partition will retain a small portion of total $\delta N$. In a special limit, $dN=\protect\underset{N\rightarrow\infty}{\lim}\delta N$, this is the context of equation (\ref{eq:2}). (c) The same as (b) in the sense of large cluster and many partitions, but we extend from $N$ which is total number of particles, to two properties $A$ and $B$ which can represents any physical quantity, in general sense $A$ or $B$ can be \{molecules number, mass, energy, temperature, etc\} in the context that one property can be exchanged to another as physical phenomenon, this concepts are relateds to equations (\ref{eq:3})-(\ref{eq:9}). }

\end{figure}


\begin{equation}
\sum_{i=1}^{^{n}}\left(\frac{N_{i}}{N}\right)=1.\label{eq:1}
\end{equation}
For simplicity we can see that a portion instead of sum of all contributions,
can be denoted as,

\begin{equation}
\left(\frac{\delta N}{N}\right)=c_{N}.\label{eq:2}
\end{equation}

From equation (\ref{eq:1}) we can see that the sum of all portions
of molecules number reach the maximum value which is the total of
molecules amount. Otherwise, a small portion of molecules number presented
in (\ref{eq:2}) gives us the fluctuation as small fraction of molecules
amounts much less than one.

Can we extend this concept of fluctuations to more than one property?
Does this extension give us some new aspects? To reach a concise response
to this question, let us relate more than one property in some balance
around frontier of a surrounded control volume.

Let us consider two different properties $A$ and $B$, and a region
in the space like a control volume or membrane, at which this two
properties can exchange one to another inside this boundaries region.
So considering this,
\begin{equation}
\left(\frac{\delta A}{A}\right)=c_{A},\label{eq:3}
\end{equation}
and
\begin{equation}
\left(\frac{\delta B}{B}\right)=c_{B}.\label{eq:4}
\end{equation}
Here $c_{A}$ and $c_{B}$ are fractions of a total amount, their
total summation reaches one. Then, summing all portions over complete
region from (\ref{eq:3}) and (\ref{eq:4}) we obtain,
\begin{equation}
\sum_{n}\left(\frac{\delta A}{A}\right)=1,\label{eq:5}
\end{equation}
and
\begin{equation}
\sum_{n}\left(\frac{\delta B}{B}\right)=1.\label{eq:6}
\end{equation}
Observing (\ref{eq:5}) and (\ref{eq:6}) we can see as first impression
a balance condition as,
\begin{equation}
\sum_{n}\left(\frac{\delta A}{A}\right)=1=\sum_{n}\left(\frac{\delta B}{B}\right).\label{eq:7}
\end{equation}

But the question doesn't close so easily, because in physical cases
we have some law relating $A$ and $B$. As a good choice we point
here to relate fraction of quantities as their appear in (\ref{eq:3})
and (\ref{eq:4}) which gives,

\begin{equation}
\left(\frac{\delta A}{A}\right)=\frac{c_{A}}{c_{B}}\left(\frac{\delta B}{B}\right),\label{eq:8}
\end{equation}
that can be simplified to initial form as,

\begin{equation}
\left(\frac{\delta A}{A}\right)\propto\left(\frac{\delta B}{B}\right).\label{eq:9}
\end{equation}

As example, our first case study starts from temperature, considering
it as important parameter of a gas, and for this reason let's analyze
the fluctuation of temperature, searching for new useful relations
for balance of properties in a gas.

\section{Boltzmann Distribution PDE}

\subsection{A Closer View to First Law of Thermodynamics }

Let our thermodynamics system with internal energy $U$, $Q$ as energy
that external agent delivery to the system and $-W$ the work done
by system. The portion $dQ$ is delivered by heating system at each
cycle and some portion of energy that we deliver to the system will
be spent with mechanical work $-dW$. Therefore, the internal energy
of system, which is a \textbf{balance} of \textbf{energy delivered}
to the system\textbf{ $dQ$} with discount of the \textbf{mechanical
work done} by the system $-dW$, is given by,

\begin{equation}
dU=dQ-dW,\label{eq:10}
\end{equation}
which is the First Law of Thermodynamics \cite{Mortimer2000}.

For simple analysis, with the aim of verify temperature fluctuations,
we can put the equation (\ref{eq:10}) in another terms: changing
the heating energy of system $dQ$ to material components temperature
increase as $CdT$, where $C$ is thermal capacity and $T$ is temperature,
and thinking as a work done $-dW$ like expansion of system. We get

\begin{equation}
dU=CdT-pdV.\label{eq:11}
\end{equation}
Isolating small fluctuation of temperature, we can see that,

\begin{equation}
\frac{dT}{T}=\frac{dU+pdV}{CT}.\label{eq:12}
\end{equation}
Equation (\ref{eq:12}) shows that temperature fluctuation is related
to internal energy modification and work done by system.

In next steps we will consider how these laws can contribute to understand
gas dynamics.

\subsection{Small Fluctuation Balance to Gas System}

As we can see at Right Hand Side (RHS) of equation (\ref{eq:12}),
the internal energy term and work term are divided by thermal energy
amount, expressing energy fluctuations and this fact explains the
fundamental relation between temperature fluctuation and energy fluctuation.
As we discuss at equation (\ref{eq:9}), a first appointment to this
energy-temperature balance of fluctuations can be done as,

\begin{equation}
\frac{\delta E}{E}\propto\frac{\delta T}{T}.\label{eq:13}
\end{equation}
This relation is based on thermal system, which has a balance between:
input energy, work done and internal energy of system. The continuation
of the analysis that improve equation (\ref{eq:13}), will be made
for a gas inside a box as our object of study and the most important
thing to study this gas dynamic system. The analysis consists of a
small virtual work\footnote{As default in Classical Mechanics to obtain motion equations.}
considering the box frontiers slightly moving from initial state to
most outside or inside position. This analysis will give us how dependency
between energy and temperature fluctuations increase or decrease.

We start this analysis first considering a gas inside a box with $T$
as gas temperature and $E$ as energy, as two fundamental quantities\footnote{In this context, temperature $T$ is a macroscopic value and energy
$E$ is a microscopic value related to the first one.}. If we think this gas can do small displacement pushing box walls
to outside, like a small virtual work, we need to balance where the
Energy fluctuation will be related to work done by gas, which will
cost a loss of thermal energy of the gas and as effect temperature
will decrease\footnote{The work done by gas costs internal energy decreasing the temperature
of a gas.}. In mathematical terms equation (\ref{eq:13}) became,

\begin{equation}
\frac{\delta E}{E}\propto-\frac{\delta T}{T}.\label{eq:14}
\end{equation}

Another change that must be done is related to dimensional analysis,
because Left Hand Side (LHS) is a fluctuation of Energy, and RHS is
related to temperature.

At first glance the thermal energy $kT$ could be a good proposition,
considering that a gas inside a box is well represented by temperature
of gas in insulated system. Then,

\begin{equation}
\frac{\delta E}{E}=-\frac{\delta kT}{kT},\label{eq:15}
\end{equation}
as we can see, a natural simplification of constant $k$ occurs. This
is a strong aspect of fluctuation analysis, because all fluctuation
terms are dimensionless in essence.

In the next steps let's consider this fractions $\delta E$ and $\delta T$
so small that we can assume them as differentials quantities $dE$
and $dT$. So,
\begin{equation}
\frac{dE}{E}=-\frac{dT}{T}.\label{eq:16}
\end{equation}
Considering that we can have some distribution function like $u\left(E,T\right)$
and other dependencies of interest, a partial differential approach
is most affordable to this purpose, then we can change (\ref{eq:16})
to,

\begin{equation}
\frac{\partial E}{\partial T}=-\frac{E}{T}.\label{eq:17}
\end{equation}
Using chain rule below, with the goal to include distribution function
$u$$\left(E,T\right)$, given by

\begin{equation}
\frac{\partial E}{\partial T}=\frac{\partial u}{\partial T}\frac{\partial E}{\partial u}.\label{eq:18}
\end{equation}
Replacing (\ref{eq:18}) in (\ref{eq:17}), we get a partial differential
equation in the form,

\begin{equation}
\frac{\partial u}{\partial T}=-\frac{E}{T}\frac{\partial u}{\partial E}.\label{eq:19}
\end{equation}
This is a new form to study distribution, from fluctuation analysis.
\textbf{Here, we present the Boltzmann Distribution PDE}. Now, to
obtain the Boltzmann distribution, we only need to solve Partial Differential
Equation (\ref{eq:19}).

We point here a function that depends on two properties which we made
fluctuation analysis. For simplicity, we use a fashion of exponents
$a$ and $b$ like we do in dimensional analysis,
\begin{equation}
u\equiv e^{-(kT)^{a}\cdot(E)^{b}}.\label{eq:20}
\end{equation}
Replacing (\ref{eq:20}) in (\ref{eq:19}), we get,
\begin{equation}
\frac{\partial u}{\partial T}=-k^{a}E^{b}aT^{a-1}e^{-(kT)^{a}\cdot(E)^{b}},\label{eq:21}
\end{equation}

\begin{equation}
\frac{\partial u}{\partial E}=-k^{a}T^{a}bE^{b-1}e^{-(kT)^{a}\cdot(E)^{b}}.\label{eq:22}
\end{equation}
To relate (\ref{eq:21}) and (\ref{eq:22}), we just divide booth
terms and simplify in order to analyse our general solution. Then,

\begin{equation}
\frac{\partial u}{\partial T}=\frac{a}{b}\frac{E}{T}\frac{\partial u}{\partial E}.\label{eq:23}
\end{equation}

The way between (\ref{eq:20}) to (\ref{eq:23}) show us that $a$
and $b$ are parameters that we can choose to satisfy the PDE. Finally,
comparing (\ref{eq:23}) to (\ref{eq:19}) we can see that choosing
$a=-b$ set PDE in (\ref{eq:23}) as the same as Boltzmann PDE in
(\ref{eq:19}), with the same sort this choice implies in tentative
solution of equation (\ref{eq:20}) to the form,

\begin{equation}
u=e^{-\left(E/kT\right)^{b}}.\label{eq:24}
\end{equation}
At last step, if we choose $b=1$, we find the Boltzmann Distribution
function\footnote{Note that we start solving a PDE using a general solution method inspired
by exponentials combinations, like when we try find some law of Physics
using Dimensional Analysis. As expected, this method gives us free
to choose some parameters to get result, but it's a strong point of
this method, not a weakness.},
\begin{equation}
u\propto e^{-E/kT}.\label{eq:25}
\end{equation}

Next step we study application of this distribution in Planck hypothesis
context, which refers to using Boltzmann Distribution function to
get Planck's Radiation Law.

\section{Planck's Law as Boltzmann Distribution PDE Solution}

\subsection{Planck hypothesis}

Planck when solving ultraviolet catastrophe explains two points. First
one is their achievement foundations related to Boltzmann Distribution
and second that he finds their formula heuristically searching for
a curve that satisfy experimental facts.

In our Appendix A session, we treat aspects related to Planck \textit{Ansatz
}which consist proposing that number of states like a photon gas is
related to Boltzmann distribution, and we point additionally to particular
regime $h\nu\ll kT$ as an experimental foundation too. Boltzmann
distribution is given by\footnote{Another solution to Boltzmann Distribution PDE (\ref{eq:19}) have
positive signal.}

\begin{equation}
u\propto e^{+E/kT}.\label{eq:26}
\end{equation}
Replacing $E=h\nu$ in (\ref{eq:26}), the initial form of distribution
stay

\begin{equation}
u=e^{h\nu/kT}.\label{eq:27}
\end{equation}
Planck solves breaking region in ultraviolet zone, which occurs because
$kT$ term is much greater than $h\nu$, or just, $kT\gg h\nu$.

At same fashion as Einstein explains easily mass-energy equivalence,
expanding kinetic energy because of velocity $v$ is much less than
$c$ \cite{Einstein2012}, let's expand Boltzmann distribution using Planck
hypothesis, with Taylor series at variable $\nu$ around ultraviolet
zone limit. So,

\begin{equation}
e^{h\nu/kT}=\lim_{h\nu\ll kT}e^{h\nu/kT}\left[1+\frac{h\nu}{kT}+\frac{1}{2}\left(\frac{h\nu}{kT}\right)^{2}+...+\sum_{n=3}\frac{1}{n!}\left(\frac{h\nu}{kT}\right)^{n}\right].\label{eq:28}
\end{equation}
Considering $h\nu\ll kT$ as in (\ref{eq:28}), let's truncate terms
to get an approximation, for all terms with $n$ equal two or higher,
resulting,

\begin{equation}
e^{h\nu/kT}\simeq1+\frac{h\nu}{kT}.\label{eq:29}
\end{equation}
Isolating $kT$ term,

\begin{equation}
kT\simeq\frac{h\nu}{e^{h\nu/kT}-1},\label{eq:30}
\end{equation}
that is a thermal energy related to a photon in the gas at state $\mu$.
Or just $E_{\mu}$,

\begin{equation}
E_{\mu}=\frac{h\nu}{e^{h\nu/kT_{\mu}}-1}.\label{eq:31}
\end{equation}
This equation is the Planck's Law of Radiation, given energy per photon
state $\mu$ in the gas, valid across ultraviolet zone. Planck shows
by his law a mathematical relation that attends phenomenology not
solved before by the classical formulation.

The essence of Planck proposition is that his \textit{Ansatz }consists
of relating radiation as gas of photons and applying Boltzmann distribution
as hypothesis to study this gas.

\subsection{Planck's Law as Boltzmann's Differential Equation Solution }

When Planck starting from Boltzmann distribution and obtain their
radiation law, he defines the case study as a photon gas system which
has compatibility with Boltzmann gas. In this perspective that Planck's
law was originated from Boltzmann Distribution, let's investigate
if Planck's law satisfies Boltzmann PDE equation (\ref{eq:19}),

\begin{equation}
\frac{\partial u}{\partial T}=-\frac{E}{T}\frac{\partial u}{\partial E},\label{eq:32}
\end{equation}
or just,

\begin{equation}
\frac{\partial u}{\partial T}\left(\frac{\partial u}{\partial E}\right)^{-1}=-\frac{E}{T}.\label{eq:33}
\end{equation}
Starting from equation (\ref{eq:31}),

\begin{equation}
E_{\mu}\equiv u=\frac{1}{e^{h\nu/kT_{\mu}}-1}.\label{eq:34}
\end{equation}
Without loss of generality let's introduce two free parameters $a$
and $b$ in (\ref{eq:34}). Then,

\begin{equation}
u=\left(e^{h\nu/kT_{\mu}}+a\right)^{b}.\label{eq:35}
\end{equation}
Applying (\ref{eq:35}) in (\ref{eq:33}), we get

\begin{equation}
\frac{\partial u}{\partial T}=-b\left(e^{h\nu/kT_{\mu}}+a\right)^{b-1}\frac{h\nu}{kT_{\mu}^{2}}e^{h\nu/kT_{\mu}},\label{eq:36}
\end{equation}
and,

\begin{equation}
\frac{\partial u}{\partial E}=b\left(e^{h\nu/kT}+a\right)^{b-1}\frac{1}{kT_{\mu}}e^{h\nu/kT_{\mu}}.\label{eq:37}
\end{equation}
Relating (\ref{eq:36}) and (\ref{eq:37}) at same fashion we relate
(\ref{eq:21}) and (\ref{eq:22}), gives

\begin{equation}
\frac{\partial u}{\partial T}=-\frac{h\nu}{T_{\mu}}\frac{\partial u}{\partial E}.\label{eq:38}
\end{equation}
So, we can see that Planck's Radiation law satisfies Boltzmann equation
with $h\nu$ as energy $E$.

Another question that we can investigate, is related to modeling thermodynamic
systems. Macroscopic variables like temperature, pressure and volume
can be measured by experimentation. Otherwise, we know that these
ones are manifestation of micro states configuration, as the case
of temperature which is the average molecular kinetic energy. The
entropy is a quantity that turns possible to relate micro states to
macro states \cite{Endrew1984}.

Planck interpreted radiation as a photon gas and obtain their law
with success. In next session let's investigate how to relate entropy
to distribution of microstates using Boltzmann PDE.

\section{Boltzmann Entropy Law}

In previous sections we have success when obtaining some known distributions.
Considering their close relationship with Statistical Physics, let
we show how to obtain Boltzmann's Entropy Law using a distribution.

Supposing some gas volume and their free expansion, when volume grow
entropy increases, but growing entropy by increasing volume costs
thermal energy of gas which will decay. So,

\begin{equation}
\frac{\delta\left(E_{S}\right)}{E_{S}}=-\frac{\delta\left(E_{T}\right)}{E_{T}}.\label{eq:39}
\end{equation}
At proper limit we can adjust this equation to,

\begin{equation}
\frac{dE_{S}}{E_{S}}=-\frac{dE_{T}}{E_{T}}.\label{eq:40}
\end{equation}
Let we set a distribution $\Omega$ which will depends on energy displaced
to entropy $E_{S}$ when volume increases and the thermal energy costs
to volume increase $E_{T}$. Given $\Omega\equiv\Omega\left(E_{S},E_{T}\right)$,
we must change from total to partial differential equation for this
purpose,

\begin{equation}
\frac{\partial E_{S}}{\partial E_{T}}=-\frac{E_{S}}{E_{T}}.\label{eq:41}
\end{equation}
Applying the chain rule like below,
\begin{equation}
\frac{\partial E_{S}}{\partial E_{T}}=\frac{\partial\Omega}{\partial E_{T}}\frac{\partial E_{S}}{\partial\Omega}.\label{eq:42}
\end{equation}
Replacing (\ref{eq:42}) in (\ref{eq:41}) we get a final form of
partial differential equation which we name as \textbf{Boltzmann's
Entropy Distribution PDE},

\begin{equation}
\frac{\partial\Omega}{\partial E_{T}}=-\frac{E_{S}}{E_{T}}\frac{\partial\Omega}{\partial E_{S}}.\label{eq:43}
\end{equation}
Our Ansatz to solve this equation consists in use the tentative solution
below,

\begin{equation}
\Omega\left(E_{S},E_{T}\right)\equiv e^{E_{S}^{a}E_{T}^{b}}.\label{eq:44}
\end{equation}
To obtain the PDE solution let we derive our tentative from (\ref{eq:44}).
First in terms of $E_{T}$ given $\frac{\partial\Omega}{\partial E_{T}}=bE_{S}^{a}E_{T}^{b-1}\Omega$
and after in terms of $E_{S}$, $\frac{\partial\Omega}{\partial E_{S}}=aE_{S}^{a-1}E_{T}^{b}\Omega$.
Relating both derivatives we can equate,
\begin{equation}
\frac{\partial\Omega}{\partial E_{T}}\left(\frac{\partial\Omega}{\partial E_{S}}\right)^{-1}=\frac{bE_{S}^{a}E_{T}^{b-1}\Omega}{aE_{S}^{a-1}E_{T}^{b}\Omega}=\frac{b}{a}\frac{E_{S}}{E_{T}}.\label{eq:45}
\end{equation}
To satisfy Boltzmann's Entropy Distribution PDE we can set $b/a=-1$,
where our particular choice is $b=-a$. So, finally we obtain the
solution of PDE in (\ref{eq:43}),

\begin{equation}
\Omega\left(E_{S},E_{T}\right)=e^{E_{S}/E_{T}}.\label{eq:46}
\end{equation}

At next steps let we do a connection between this distribution to
Boltzmann's Entropy law. Using as first principles dimensional analysis
and thermodynamics laws. From entropy fundamental concept defined
in Thermodynamics $dQ=TdS$ from who we can see that $\left[dQ\right]=dE_{S}$
establishing relation between heat energy and entropy energy, and
as macroscopic version we can set dimensional dependence of entropy
energy as $E_{S}\propto\left[T\right]^{a}\left[S\right]^{b}$ which
dimensional solution points to $E_{S}=TS$. Another demand is to thermal
energy $E_{T}$ which we get as $E_{T}=kT$. So,

\begin{equation}
\Omega\left(E_{S},E_{T}\right)=e^{TS/kT}=e^{S/k},\label{eq:47}
\end{equation}
where, to isolate entropy $S$ in the final form, let we assume the
inverse function from (\ref{eq:47}). Wich gives

\begin{equation}
S=k\ln\left|\Omega\right|.\label{eq:48}
\end{equation}

\section{Fermi-Dirac and Bose-Einstein Distributions}

Now, we will verify how to Fermi-Dirac and Bose-Einstein are related
to a similar format as that obtained in equation of Planck's law (\ref{eq:35}).
So,

\begin{equation}
u=\left(e^{E/kT}+a\right)^{b}.\label{eq:49}
\end{equation}
Here, our emphasis consists in using the Boltzmann's partial differential
equation to inspect how close Fermi-Dirac and Bose-Einstein are to
Plank's law format. With a little modification in energy from (\ref{eq:49}),
we get

\begin{equation}
u=\left(e^{\left(\varepsilon-\mu\right)/kT}+a\right)^{b}.\label{eq:50}
\end{equation}
 And the analogous form to (\ref{eq:33}), is

\begin{equation}
\frac{\partial u}{\partial T}\left(\frac{\partial u}{\partial\varepsilon}\right)^{-1}=-\frac{\left(\varepsilon-\mu\right)}{T}.\label{eq:51}
\end{equation}
Next steps let's apply (\ref{eq:50}) in (\ref{eq:51}) to study both
Fermi-Dirac and Bose-Einstein particular cases.

\subsection{Fermi-Dirac and Bose-Einstein Distributions Analysis}

Comparing general format from (\ref{eq:50}) with Fermi-Dirac distribution
equation in \cite{Crawford1988} we see that $a$ and $b$ must assume,

\begin{equation}
\left(a,b\right)\equiv\left(+1,-1\right).\label{eq:52}
\end{equation}
This choice gives us,

\begin{equation}
u=\frac{1}{e^{\left(\varepsilon-\mu\right)/kT}+1}.\label{eq:53}
\end{equation}
Derivating in temperature and energy we have their respective equations,

\begin{equation}
\frac{\partial u}{\partial T}=\left[e^{\left(\varepsilon-\mu\right)/kT}+1\right]^{-2}e^{\left(\varepsilon-\mu\right)/kT}\frac{\left(\varepsilon-\mu\right)}{kT^{2}},\label{eq:54}
\end{equation}
and

\begin{equation}
\frac{\partial u}{\partial\varepsilon}=-\left[e^{\left(\varepsilon-\mu\right)/kT}+1\right]^{-2}e^{\left(\varepsilon-\mu\right)/kT}\frac{1}{kT}.\label{eq:55}
\end{equation}
Now relating both (\ref{eq:54}) and (\ref{eq:55}), we finally see
that,

\begin{equation}
\frac{\partial u}{\partial T}\left(\frac{\partial u}{\partial\varepsilon}\right)^{-1}=-\frac{\left(\varepsilon-\mu\right)}{T},\label{eq:56}
\end{equation}
Fermi-Dirac's distribution satisfy general form of Boltzmann's distribution
PDE. At same fashion, comparing general format from (\ref{eq:50})
with Bose-Einstein distibution equation in \cite{Barford1976} we see
that $a$ and $b$ must assume,

\begin{equation}
\left(a,b\right)\equiv\left(-1,-1\right).\label{eq:57}
\end{equation}
This choice gives us,

\begin{equation}
u=\frac{1}{e^{\left(\varepsilon-\mu\right)/kT}-1}.\label{eq:58}
\end{equation}
Which verification results will led us to, at same way that obtained
in (\ref{eq:56}), concluding that Bose-Einstein's distribution satisfy
general form of Boltzmann's distribution PDE too.

So, when we treat Plank's law of radiation, we see that same PDE which
gives us Boltzmann's Distribution is satisfied to Planck's law. We
extend this analysis using a general form of distribution and conclude
that also Fermi-Dirac's and Bose-Einstein's distributions emerges
as Planck's law from the same Boltzmann's PDE.

\section{Schrödinger's Distribution}

The uncertainty principle emerges from experimental and Gaussian widths
relations \cite{Nairz2002}. We use uncertainty as strong fundamental
to point our dimensionless fluctuation and obtain a fundamental distribution
that we can use to recover Schrödinger equation.

To obtain Schrödinger distribution, we must recover the most
fundamental aspect, which lies over uncertainty relations \cite{Heisenberg1927},

\begin{equation}
\delta x\delta p\geq\frac{\hbar}{2},\label{eq:59}
\end{equation}
and

\begin{equation}
\delta E\delta t\geq\frac{\hbar}{2}.\label{eq:60}
\end{equation}
Another association that we can make consist in show interrelation
between booth (\ref{eq:59}) and (\ref{eq:60}), starting from energy
relation $\delta E=m\delta(v)^{2}/2=mv\delta v$. Replacing it in
(\ref{eq:60}) we get $v\delta t\cdot m\delta v\geq\hbar/2$, and
remember that $\delta x=v\delta t$ and $\delta p=m\delta v$ we recover
the other uncertainty relation $\delta x\delta p\geq\hbar/2$ as in
refference \cite{Heisenberg1927}.

Now, we are supposing a system with some initial energy amount $E$
and a free particle with momentum $p$ and an isolated system. In
terms of dimensionless fluctuation analysis, if the particle momentum
$p$ increase a quantity $\delta p$, it will cost to system a decrease
amount of energy $-\delta E$. So, we can write

\begin{equation}
\frac{\delta p}{p}\propto-\frac{\delta E}{E}.\label{eq:61}
\end{equation}

As next steps, we start replacing uncertainty relations from (\ref{eq:59})
and (\ref{eq:60}), in the limit that we equate both. So,

\begin{equation}
\delta E\delta t=\delta x\delta p.\label{eq:62}
\end{equation}
Multiplying (\ref{eq:62}) by inverse of $E$ and, after a little
algebra, we get

\begin{equation}
\frac{\delta E}{E}=2\frac{\delta p}{p}.\label{eq:63}
\end{equation}

Now comparing this format of uncertainty relation presented in (\ref{eq:63})
to dimensionless fluctuation relation of (\ref{eq:61}), we can see
that instead of proportionality factor assume value $1$ as in many
physical cases \cite{Barenblat1987}, a factor $2$ will be needed, given

\begin{equation}
\frac{\delta E}{E}=-2\frac{\delta p}{p}.\label{eq:64}
\end{equation}
After a little algebra, and considering the proper limit when $\delta E$
and $\delta p$ can be assumed as differentials, we get

\begin{equation}
\frac{dE}{dp}=-2\frac{E}{p}.\label{eq:65}
\end{equation}
Assuming that we have some distribution function, $\Psi\equiv\Psi\left(p,E\right)$,
the partial derivatives turn (\ref{eq:65}) to a most interesting
format,

\begin{equation}
\frac{\partial E}{\partial p}=-v.\label{eq:66}
\end{equation}
Assuming a chain rule using $\Psi$, relating $E$ and $p$ we get,
\begin{equation}
\frac{\partial E}{\partial p}=\frac{\partial\Psi}{\partial p}\frac{\partial E}{\partial\Psi}.\label{eq:67}
\end{equation}
Replacing (\ref{eq:67}) in (\ref{eq:66}), we finally get the \textbf{Schrödinger
distribution equation PDE},

\begin{equation}
\frac{\partial\Psi}{\partial p}=-v\frac{\partial\Psi}{\partial E}.\label{eq:68}
\end{equation}
We can easily see that a simple solution is given by,

\begin{equation}
\Psi=e^{ipx/\hbar}e^{-iEt/\hbar}.\label{eq:69}
\end{equation}
This distribution equation is very important, because from this one
we can obtain all properties of the system, including schrödinger
Equation. From expected value as statistical concept $\left\langle p\right\rangle =\int_{-\infty}^{+\infty}p\Psi\left(x,t\right)dx$,
we can see that distribution play a central role. Because of complex
variable format at (\ref{eq:69}) we extend this to square integrate
as $\left\langle p\right\rangle =\int_{-\infty}^{+\infty}\Psi^{*}\left(x,t\right)\hat{p}\Psi\left(x,t\right)dx$.
At same fashion for $\left\langle H\right\rangle $ and $\left\langle V\right\rangle $
and after putting it all together in Hamiltonian energy conservative
form of square integrable expected values, we easily recover Schrödinger
equation.

\section{Concluding Remarks}

After our tests results was verified with success to principal Physics
Distributions, we find a collection of procedures as registered in
this paper, which we can summarize in a single principle: ``\textit{Given
some fundamental relations between two or more quantities, the Dimensionless
Fluctuation Analysis point to a Partial Differential Equation which
solution gives desired distribution}''. Which we name as\textbf{
Dimensionless Fluctuation Balance Principle.} At least this work shows
a new way to relate two quantities using dimensionless number that
enable us to obtain distributions as differential equations solutions.
This is not only a practical tool, but also turns possible to understand
some meaning beyond distributions. As example, we can see that when
some particle or energy or another system entity was confined submitted
to some compact region like localized phenomena or when required to
interact with some other entities not in large space field, distributions
emerges as natural imposition of space region to a group of entities
when the system size is so small as needed to impose statistical reality
to every entity within it.

\section{Appendix A}

Observing black line in the figure (\ref{fig:2}), winch is a curve
for classical theory to $5000K$, this curve diverges when compared
with Planckian blue curve to $5000K$.

In classical curve, a breaking point appears in classical curve near
$\lambda=1\times10^{-6}m$ but we can observe this problem starts
at the utraviolet region, because other curves like red $3000K$,
green $4000K$ and blue $5000K$, which is expected by experiments,
cross ultraviolet region without blow up.
\noindent \begin{center}
\begin{figure}[ht]
\noindent \begin{centering}
\includegraphics[scale=0.60]{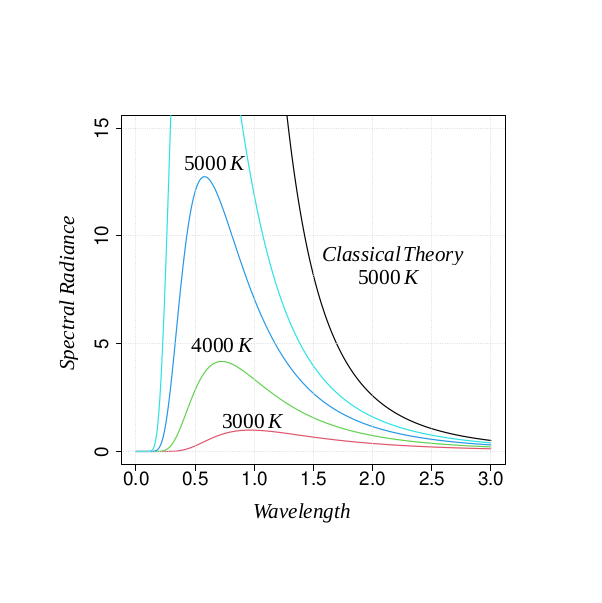}
\par\end{centering}
\caption{
\label{fig:2}Classical radiation distribution, evidencing temperature
of $5000K$ and a breaking point near wavelength $1\mu m$. Graphic
units: wavelenght $\mu m$ and Spectral Radiance $kW\cdot sr^{-1}\cdot m^{-2}\cdot nm^{-1}$.}
\end{figure}
\par\end{center}

In other words we need to investigate $\lambda=0.375\times10^{-6}m$,
which is the ultraviolet region, where the classical curve problem
begin.

Considering $c=\lambda\nu$, we easily get $\nu=8\times10^{14}Hz$
as frequency in the point that wavelength is $\lambda=0.375\times10^{-6}m$.

To understand this limit when classical curve blow up at ultraviolet
region, let's investigate the relation $\left(h\nu/kT\right)$, which
appears in equation (\ref{eq:28}), using ultraviolet frequency $\nu=8\times10^{14}Hz$
and $T=5\times10^{3}K$ as in figure (\ref{fig:2}). Then,
\begin{equation}
\frac{h\nu}{kT}=\frac{\left(6.63\times10^{-34}m^{2}\cdot Kg\cdot s^{-1}\right)\left(8\times10^{14}s^{-1}\right)}{\left(1.38\times10^{-23}m^{2}\cdot Kg\cdot s^{-2}\cdot K^{-1}\right)\left(5\times10^{3}K\right)}=0.77\times10^{-43}.\label{eq:70}
\end{equation}

Equation (\ref{eq:41}) tell us that breaking region in ultraviolet
zone occurs because $kT$ term is much greater than $h\nu$. Or just,

\begin{equation}
h\nu\ll kT.\label{eq:71}
\end{equation}

\section{Appendix B}

In this session, we just recover Schrödinger equation in a quick way,
starting from \textbf{Schrödinger distribution obtained as PDE solution}
(\ref{eq:69}) as a fundamental to discover operators from mean value
concept, the aim of this session is show a close relation between
a distribution that became from experimental measurable parameters
like uncertainty relations and schorödinger quantum mechanics.

Starting from Hamiltonian of expected values $\left\langle H\right\rangle =\left\langle p\right\rangle ^{2}/2m+\left\langle V\right\rangle $
in square integrable system, we easily can see that,

\begin{equation}
\int_{-\infty}^{+\infty}\Psi^{*}\left(x,t\right)\hat{H}\Psi\left(x,t\right)dx=\frac{1}{2m}\int_{-\infty}^{+\infty}\Psi^{*}\left(x,t\right)\hat{p}^{2}\Psi\left(x,t\right)dx+\int_{-\infty}^{+\infty}\Psi^{*}\left(x,t\right)\hat{V}\Psi\left(x,t\right)dx.\label{eq:72}
\end{equation}
Another useful concept is that $\hat{p}$ operator must extract eigenvalue
from distribution, as in the form $\hat{p}^{2}\Psi\left(x,t\right)=p^{2}\Psi\left(x,t\right)$.
Because of format of distribution we get in (\ref{eq:69}) only a
specific $\hat{p}$ operator can extract $p$ eigenvalue. So,

\begin{equation}
\Psi\left(x,t\right)=e^{ipx/\hbar}e^{-iEt/\hbar}\Leftrightarrow\hat{p}\equiv-i\hbar\frac{\partial}{\partial x}.\label{eq:73}
\end{equation}
At same fashion, we get $\hat{H}$ operator (that extract energy $E$)
and $\hat{V}$ operator (that impose potential energy $V$) in (\ref{eq:72})
to get the Schrödinger equation from expected values and Schrödinger
distribution. Which gives,

\begin{equation}
\int_{-\infty}^{+\infty}\Psi^{*}\left(x,t\right)\left[i\hbar\frac{\partial}{\partial t}+\frac{\hbar}{2m}^{2}\frac{\partial{{}^2}}{\partial x{{}^2}}-V\left(x,t\right)\right]\Psi\left(x,t\right)dx=0,\label{eq:74}
\end{equation}
or just,

\begin{equation}
i\hbar\frac{\partial}{\partial t}\Psi\left(x,t\right)=-\frac{\hbar}{2m}^{2}\frac{\partial{{}^2}}{\partial x{{}^2}}\Psi\left(x,t\right)+V\left(x,t\right)\Psi\left(x,t\right).\label{eq:75}
\end{equation}
Here, we present the \textbf{Schrödinger equation as Heisenberg uncertainty
relations Mechanics}. The strong aspect relies in distribution, because
from this one operators assume their form. We show by unique way with
fluctuations analysis that Heisenberg uncertainty principle is determinant
to obtain distribution that will lead us to find Schrödinger equation
as a simple consequence.

\section{Authors Declaration Interests}
The authors declare that they have no know competing financial interests or personal relationship that could have appeared to influence the work reported in this paper.



\begin{thebibliography}{23}

\bibitem[1]{Maxwell1864}J. Clerk Maxwell, A Dynamical Theory of the
Electromagnetic Field, Royal Society Publishing, (1864).\\
 \url{https://doi.org/10.1098/rstl.1865.0008}%


\bibitem[2]{Whestphal2021}T. Whestphal, H. Hepach, J. Pfaff and M. Aspelmeyer,
``Measurement of Gravitational Coupling Between Millimetre-Sized
Masses'', Nature Physics, (2021).\\
 \url{https://doi.org/10.1038/s41586-021-03250-7}%


\bibitem[3]{Merlin2021}R. Merlin, Rabi oscillations, Floquet states,
Fermis Golden Rule, and all that Insights From an Exactly Solvable
Two-Level Model, American Journal of Physics, 89, 26, (2021).\\
\url{https://doi.org/10.1119/10.0001897}%


\bibitem[4]{Landau1980}L D Landau, E M Lifshitz, The Classical Theory
of Fields,Elsevier,(1980).%


\bibitem[5]{Basil2011}Basil S. Davis and Lev Kaplan, Poynting Vector
Flow in a Circular Circuit, American Journal of Physics 79, 1155
(2011).\\
 \url{https://doi.org/10.1119/1.3630927}%


\bibitem[6]{Arfken2012}George B. Arfken , Hans J. Weber, et al, Mathematical
Methods for Physicists, 7th edition, Elsevier, (2012).

\bibitem[7]{Reynolds1902}Reynolds Osborne, On the sub-mechanics of the
Universe, Proc. R. Soc. Lond. 69425--433, (1902).\\
 \url{http://doi.org/10.1098/rspl.1901.0127}%


\bibitem[8]{Abayev2006}Ilana Abayev, Properties of the Electronic Density
of States in TiO2 Nanoparticles Surrounded with Aqueous Electrolyte,
J Solid State Electrochem, (2006).\\
\url{https://doi.org/10.1007/s10008-006-0220-1}%


\bibitem[9]{Pierantoni2021}Luca Pierantoni et all, Dirac Equation Based
Formulation for the Quantum Conductivity in 2D Nanomaterials, Applied
Sciences MDPI, (2021).\\
\url{https://doi.org/10.3390/app11052398}%


\bibitem[10]{Vollath2019}Dieter Vollath, Energy Distribution in an
Ensemble of Nanoparticles and its Consequences, Beilstein Journal
of Nanotechnology, (2019).\\
\url{https://doi.org/10.3762/bjnano.10.143}%


\bibitem[11]{Rowlinson2005}Rowlinson, J. S. The Maxwell-Boltzmann distribution,
Molecular Physics, 103.21-23: 2821-2828, (2005).\\
\url{http://dx.doi.org/10.1080/002068970500044749}%


\bibitem[12]{Agudelo2010}Agudelo, Andrés; Cortés, Cristóbal. Thermal radiation
and the second law. Energy, 35.2: 679-691, (2010).

\bibitem[13]{Rayleigh1900}Rayleigh, Lord. Remarks Upon the Law of Complete
Radiation''. Philosophical Magazine, 49: 539-540, (1900). \\
\url{http://dx.doi.org/10.1080/14786440009463878}%


\bibitem[14]{Trevena2010}{[}14{]} D.H. Trevena, The Boltzmann Distribution
and Related Topics, Statistical Mechanics, Woodhead Publishing,
Pages 18-26, (2010).\\
 \url{https://doi.org/10.1533/9780857099662.18}%


\bibitem[15]{Marion1980}Jerry B. Marion, Mark A. Heald, Classical
Electromagnetic Radiation, 2nd Edition, Academic Press, Pages 104-130,(1980).\\
\url{https://doi.org/10.1016/B978-0-12-472257-6.50008-2}%


\bibitem[16]{Mortimer2000}Mortimer, R. G., Work, Heat, and Energy.
Physical Chemistry, 45--93,(2000).\\
 \url{https://doi.org/10.1016/b978-012508345-4/50006-x}%


\bibitem[17]{Einstein2012}Einstein, Relativity, the Special and the
General Theory, General Press, New Delhi , 61-63 (2012).%


\bibitem[18]{Endrew1984}Keith Endrew, Entropy, American Journal
of Physics 52, 492, (1984).\\
 \url{https://doi.org/10.1119/1.13892}%


\bibitem[19]{Crawford1988}Frank S. Crawford, Using Einstein\textquoteright s
Method to Derive Both the Planck and Fermi-Dirac Distributions,
American Journal of Physics 56, 883, (1988).\\
 \url{https://doi.org/10.1119/1.15402}%


\bibitem[20]{Barford1976} W. C. Barford, Derivation of Classical and
Quantum Statistical Distributions, American Journal of Physics
44, 940, (1976).\\
\url{https://doi.org/10.1119/1.10236}%


\bibitem[21]{Nairz2002}Nairz, Olaf and Arndt, Markus and Zeilinger,
Anton, Experimental verification of the Heisenberg uncertainty
principle for fullerene molecules, Phys. Rev. A, 65, 3, pages 032109,
(2002)\\
 \url{https://doi.org/10.1103/PhysRevA.65.032109}%


\bibitem[22]{Heisenberg1927}Heisenberg, W. Über den Anschaulichen Inhalt
der Quantentheoretischen Kinematik und Mechanik. Z. Physik 43, 172--198
(1927). \\
\url{https://doi.org/10.1007/BF01397280}%

\bibitem[23]{Barenblat1987}Barenblat, G. I.,Dimensional Analysis.
USSR Academy of Science, Moscow, Gordon and Breach Science Publishers,
(1987). %


\end{thebibliography}
\end{document}